\documentclass[pdflatex,sn-mathphys-num]{sn-jnl}% Math and Physical Sciences Author Year Reference Style
\usepackage[utf8]{inputenc}
\usepackage{graphicx}%
\usepackage{multirow}%
\usepackage{amsmath,amssymb,amsfonts}%
\usepackage{amsthm}%
\usepackage{mathrsfs}%
\usepackage[title]{appendix}%
\usepackage{xcolor}%
\usepackage{textcomp}%
\usepackage{subcaption} 
\usepackage{manyfoot}%
 \usepackage{subcaption}                                                              
\usepackage{caption}
\usepackage{booktabs}%
\usepackage{algorithm}%
\usepackage{algorithmicx}%
\usepackage{algpseudocode}%
\usepackage{listings}%
\definecolor{redc}{rgb}{1, 0, 0}

\theoremstyle{thmstyleone}%
\theoremstyle{thmstyletwo}%

\theoremstyle{thmstylethree}%

\begin{document}

% \title[Towards FAIR AI: A Survey of Trends and Knowledge Graph-Enhanced Bias Mitigation]{Towards FAIR AI: A Survey of Trends and Knowledge Graph-Enhanced Bias Mitigation}

% \title{FAIR AI ATLAS: Empirical Analysis, Systematic Review, and Knowledge Graph Applications in AI Bias Research}
\title{Whose fairness? Structural concentration in AI bias research
% removed foundations
% Structural inequality in AI bias research: a bibliometric and semantic analysis
}


\author*[1]{\fnm{Abhash} \sur{Shrestha}}\email{abhash.shrestha@cair-nepal.org}
\author[1]{\fnm{Subigya} \sur{Gautam}}\email{subigya.gautam@cair-nepal.org}
\author[1]{\fnm{Anu} \sur{Sapkota}}\email{anu.sapkota@cair-nepal.org}
\author[2,3]{\fnm{Sanju} \sur{Tiwari}}\email{tiwarisanju18@ieee.org}
\author*[1,4]{\fnm{Tek Raj} \sur{Chhetri}}\email{tekraj.chhetri@cair-nepal.org;tekraj@mit.edu}
\affil*[1]{\orgname{Center for Artificial Intelligence (AI) Research Nepal}, \orgaddress{\street{Sundarharaincha-09}, \state{Koshi}, \country{Nepal}}}
% \city{City}, \postcode{100190}
\affil[2]{\orgname{Sharda University}, \orgaddress{\city{Delhi-NCR},\country{India}}}

\affil[3]{\orgname{Shodhguru Innovation and Research Labs}, \country{India}}

\affil[4]{\orgdiv{McGovern Institute for Brain Research}, \orgname{Massachusetts Institute of Technology}, \orgaddress{\street{43 Vassar Street}, \city{Cambridge}, \postcode{02139}, \state{MA}, \country{United States}}}

\maketitle

% ============================================================
% ABSTRACT (100-150 words, unreferenced)
% ============================================================
\begin{abstract}
\noindent 
Artificial intelligence increasingly mediates consequential decisions in healthcare, law, and public services, and the field has responded with an extensive methodology for measuring and mitigating bias. Yet the fairness definitions, benchmarks, and debiasing frameworks on which this methodology rests are treated as universal while being produced by a research community whose composition has never been characterized. We show that the AI bias research are structurally concentrated, and that this concentration is greatest, geographically, in precisely the domain the rest of the field inherits from. Analyzing 692 publications spanning five thematic domains, combining bibliometric analysis with semantic clustering, we find that research activity is dominated by a small set of countries, institutions, and authors, with the United States leading publication output and collaboration networks across every domain and most strongly in general fairness and bias mitigation, the largest, most-cited domain with meaningful representation across all four semantic clusters. Low- and middle-income countries remain largely absent from the community and its collaboration networks, and citation influence is highly skewed (median = 9; mean =93.5 ), indicating that a small fraction of publications disproportionately shapes the field. Because the general-fairness domain supplies the definitions and benchmarks that application areas apply, concentration of research effort in this foundational domain propagates across AI bias research as a whole - raising the concern that mitigation methods developed and validated within a narrow set of contexts may not generalize to all populations and settings where AI is deployed. We provide an interactive atlas for continuous monitoring of the field's structure.

\end{abstract}

% ============================================================
% INTRODUCTION (no heading per Nature MI format)
% ============================================================

\section{Introduction} 
Artificial intelligence (AI) is increasingly reshaping how knowledge is produced, services are delivered, and decisions are made across society. In this article, we use the term AI broadly to refer to computational paradigms such as machine learning~(ML), deep learning~(DL), and large language models~(LLMs). The rapid adoption of these systems across domains such as healthcare~\cite{AHMEDTAHA2024111307,GANGWAL2024103992} and education to governance and public services~\cite{dressel_accuracy_2018} has intensified concerns about AI bias~\citep{ferrara_fairness_2023,ntoutsi_bias_2020,9772748,Zack2024,lin-etal-2025-investigating,pandey2026dualmetricevaluationsocialbias}. We define AI bias as systematic error or variation in the behavior or outputs of an AI system that produces distorted or harmful outcomes for particular individuals or groups, often reflecting biases in data, model design, human decision-making, or broader sociotechnical structures.  Such bias can reproduce and amplify existing inequalities, with direct consequences for people’s lives when AI systems are embedded in high-stakes sociotechnical contexts. The stakes are well documented: medical AI systems overassociate Hispanic and Asian patients with tuberculosis and under-recommend computed tomography (CT) and magnetic resonance imaging (MRI) imaging for Black patients~\cite{Zack2024}, while risk-assessment tools such as COMPAS~(Correctional Offender Management Profiling for Alternative Sanctions) overestimate recidivism risk for Black individuals relative to White individuals with comparable profiles~\cite{dressel_accuracy_2018, Engel2024}. Understanding how these risks emerge, evolve, and can be mitigated has therefore become a pressing scientific and societal challenge.

Although contemporary concerns about AI bias build on decades of work on fairness, discrimination, and algorithmic decision-making, the issue has become increasingly consequential as AI systems are deployed at scale in decisions that shape people’s opportunities, rights, and access to services~\cite{10.1145/3287560.3287600, ntoutsi_bias_2020, Mehrabi2021, Caton2024}. This has prompted sustained research across domains including healthcare, criminal justice, recommender systems, and LLMs~\cite{ferrara_fairness_2023, ntoutsi_bias_2020, Mehrabi2021, hort_bias_2023, pessach2022review, Tang2023, Caton2024, Li2023, Gallegos2024, Chu2024, dudy_unequal_2025}. Despite substantial advances in characterizing AI bias and developing mitigation strategies, the problem remains unresolved. LLMs, in particular, have sharpened the difficulty: their scale, opacity, and dependence on vast training corpora make bias hard to detect, attribute, and mitigate. These technical challenges are compounded by global disparities in data, computational resources, technical capacity, and institutional support—disparities that determine who is able to study, evaluate, and influence how AI systems are built~\cite{pandey2026dualmetricevaluationsocialbias}. In addition, bias is not embedded in data and models/algorithms alone; it is also shaped by the regional, cultural, linguistic, and sociopolitical contexts in which AI systems are developed, evaluated, and deployed. A fairness criterion/norms that appears neutral in one setting may carry a different meaning or cause different harm in another, so that systems judged equitable in one context can prove biased when transferred across linguistic or cultural boundaries. For example, direct eye contact is often associated with confidence or positive engagement in many Western contexts~\citep{Pang2024}, but is perceived as disrespectful in some East Asian contexts, including Japan~\citep{10.1371/journal.pone.0118094}. These examples show that fairness is not culturally neutral: a criterion that appears equitable under one set of cultural, linguistic, or institutional assumptions can fail to generalize—or cause harm—when applied to other populations and deployment settings. This raises a question about the research landscape itself, since the regional and institutional composition of the community producing fairness frameworks shapes which contexts are represented and which definitions of fairness gain influence.

Yet the research landscape that produces this knowledge remains poorly characterized. Prior work has advanced definitions, benchmarks, and mitigation strategies, but the structure of the field itself---who produces this research, where, and in collaboration with whom---has received limited empirical scrutiny. Disparities across regions, institutions, collaborations, and application domains are frequently acknowledged~\cite{chan2021limitsglobalinclusionai} yet rarely measured. This gap is consequential because the composition of a research community shapes the knowledge it produces. Communities that dominate a field influence which harms are treated as central, which populations are represented, and which benchmarks, demographic categories, and fairness definitions become widely adopted. Citation and collaboration networks can reinforce existing centers of activity through cumulative advantage and preferential attachment, echoing the Matthew effect in science, in which recognition accrues disproportionately to already prominent actors and ideas~\cite{10.1098/rsif.2014.0378, doi:10.1126/science.159.3810.56}. Evidence from adjacent areas suggests that such concentration can shape AI research, in AI life science research, Asia leads in total publications, whereas North America and Europe contribute disproportionately to work published in high-ranking outlets and receiving higher citation impact, and international collaboration has stagnated relative to national research efforts~\citep{Schmallenbach2024}. When research concentrates in this way, mitigation strategies developed and validated within a narrow set of contexts may be presumed to generalize beyond the populations and sociocultural settings in which they were tested. Our findings in the Nepali cultural context illustrate this risk, showing that LLM bias can manifest differently in underrepresented sociocultural settings and may not be adequately captured by evaluation frameworks developed elsewhere~\cite{pandey2026dualmetricevaluationsocialbias}. 
Whether similar structural imbalances characterize AI bias research, however, remains empirically unresolved. Related evidence has begun to emerge in narrower areas. For example, Alberto et al.~\cite{Alberto2024} conducted a scientometric analysis of health AI fairness research and found that 82.2\% of authors were from high-income countries, with papers from wealthier nations receiving disproportionately more citations. However, existing work has largely examined AI bias within specific application domains, leaving open the question of whether the broader field is unevenly structured across countries, institutions, co-authorship networks, collaboration patterns, citation influence, and thematic areas. Importantly, these studies~\citep{Schmallenbach2024, Alberto2024} do not consider cross-cutting, domain-agnostic areas of research, such as general fairness and bias mitigation or graph-based fairness, in which fairness definitions, benchmarks, and debiasing frameworks are developed and then adopted across domains.

Here we examine who produces AI bias research and what the structure of that community implies both for the field's methodological foundations, such as the fairness definitions, and mitigation methods that domains share, and for the application-specific priorities of areas such as health AI, LLMs and NLP, recommender systems, and graph-based fairness. We characterize the AI bias research landscape through bibliometric and semantic analyses of 692 publications spanning five thematic domains: general fairness and bias mitigation, health and clinical AI, large language models and NLP, recommender systems, and graph-based fairness and bias mitigation. The corpus was constructed by systematically querying major databases, including IEEE Xplore, ACM Digital Library, Scopus, ScienceDirect, and Engineering Village, together with the ACM Conference on Fairness, Accountability, and Transparency proceedings, and was supplemented by citation snowballing and manual relevance screening. We examine the field’s geographic composition, institutional concentration, collaboration structure, citation dynamics, temporal evolution, and semantic organization. To distinguish broad participation from intellectual leadership, we quantify national and institutional contributions using both all-author and first-author counts. We also reconstruct a co-authorship network of 2,502 authors to assess regional concentration and within- versus cross-region collaboration, and compare citation patterns across within-corpus and global influence. Semantic clustering identifies general fairness and bias mitigation as a central domain: it is the largest and most highly cited domain, and its vocabulary cuts across the others, suggesting that it supplies many of the definitions, benchmarks, and mitigation frameworks subsequently adopted across application areas. We find that this central domain is also the most structurally concentrated, indicating that the production of broadly used fairness concepts is shaped by a relatively narrow set of countries, institutions, and collaboration networks. Together, these analyses move the question of who shapes AI bias research from assumption to measurement, and connect the structure of the research community to a methodological risk: fairness definitions and mitigation strategies developed within concentrated research settings may be adopted across domains as if they were universally applicable.

% Here, we characterize the AI bias research landscape through bibliometric and semantic analyses of 692 publications spanning five thematic domains: health and clinical AI, general fairness and bias mitigation, graph-based fairness and bias mitigation, large language models and NLP, and recommender systems. The corpus was constructed through systematically querying major databases~(IEEE Xplore, ACM Digital Library, Scopus, ScienceDirect, and Engineering Village) and the ACM Conference on Fairness, Accountability, and Transparency (FAccT) proceedings, supplemented by citation snowballing, and filtered through manual screening for relevance. We examine the field’s geographic composition, institutional concentration, collaboration structure, citation dynamics, temporal evolution, and semantic organization. To distinguish participation from leadership, we quantify national and institutional contributions using both all-author and first-author counts. We reconstruct a co-authorship network of 2{,}621 authors to assess regional concentration and within- versus cross-region collaboration, and compare citation patterns across within-corpus and global influence. Together, these analyses move the question of who shapes AI bias research from assumption to measurement, revealing structural inequalities in the production and circulation of knowledge on AI bias.

% ============================================================
% RESULTS
% ============================================================
\section{Results}

\subsection*{Domain distribution and temporal trends}

\begin{figure*}[htb]                        
  \centering                              
      \begin{subfigure}[t]{1\textwidth}                             \centering                                                   \includegraphics[width=\textwidth]{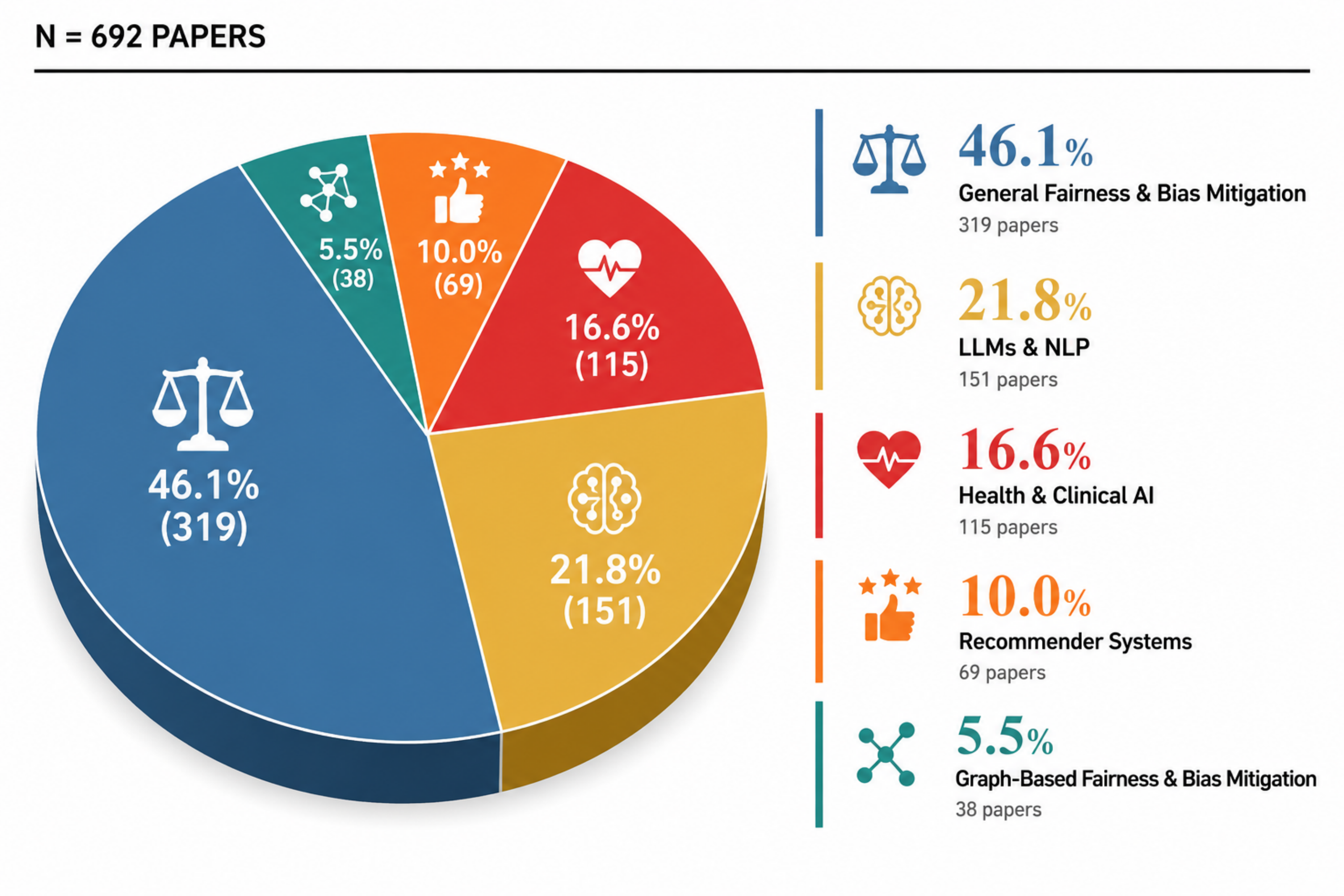}                    \caption{\textbf{(a)}}                                \label{fig:domain}                                             \end{subfigure}                                                 \hfill                                                          \begin{subfigure}[t]{1\textwidth}
          \centering                                                  \includegraphics[width=\textwidth]{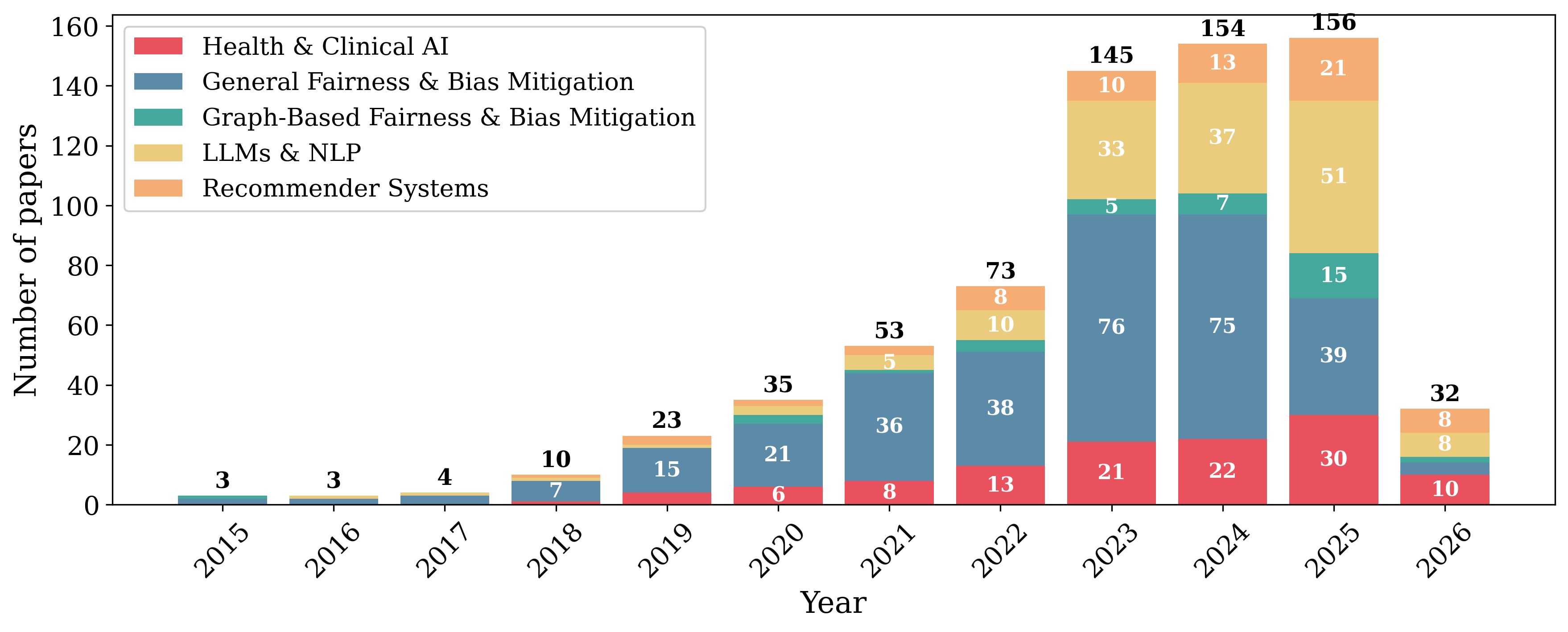}                            \caption{\textbf{(b)}}
          \label{fig:temporal}                                        \end{subfigure}                         
\caption{Domain distribution and temporal dynamics. \textbf{(a)} Distribution of 692 papers across five thematic domains. \textbf{(b)} Annual publication volume by domain (2015--2026). The post-2022 surge is  
driven by LLMs \& NLP and General Fairness \& Bias Mitigation.}
  \label{fig:domain-temporal}                                         
  \end{figure*}

Each of the 692 papers in the corpus was manually assigned to one of five thematic domains based on its primary research focus (Fig.~\ref{fig:domain}): 
General Fairness \& Bias Mitigation
(46.1\%, 319 papers), LLMs \& NLP (21.8\%, 151 papers), Health \& Clinical AI (16.6\%, 115 papers), Recommender Systems (10.0\%, 69 papers), and  
Graph-Based Fairness \& Bias Mitigation (5.5\%, 38 papers).
The predominance of General Fairness suggests the field's continued emphasis on foundational, domain-agnostic debiasing methods, frameworks that export into application-specific contexts rather than addressing them directly. 
Graph-based methods, despite their structural advantages for encoding relational knowledge, remain the least explored domain by a wide margin.                    

Publication output grew gradually from 2015 through 2021 (53 papers), before a sharp increase in 2022 - 2023 (Fig.~\ref{fig:temporal}). Annual output reached  145 papers in 2023, rose further to 154 in 2024, and reached 156 papers in 2025, roughly a three-fold increase compared with 2021. 
The lower count for 2026 (32 papers) reflects an incomplete year at the time of corpus construction rather than a decline in research activity.
This surge was not uniform across domains. General Fairness \& Bias Mitigation and LLMs \& NLP drive the post-2022 surge, coinciding with the widespread adoption of large language models. Health \& Clinical AI maintained consistent output throughout the period, suggesting that clinical AI fairness research is less sensitive to generative model developments.
The divergence in growth rate across domains suggests that publication activity is affected by domain-specific events and developments.

\subsection*{Country-level contribution and collaboration patterns}

% Page 1: subfigures (a) and (b) — no main caption
\begin{figure*}[p]
  \centering
  \hspace{-8em}
  \begin{subfigure}[t]{1\textwidth}
  
      \centering
      \includegraphics[width=1.2\textwidth]{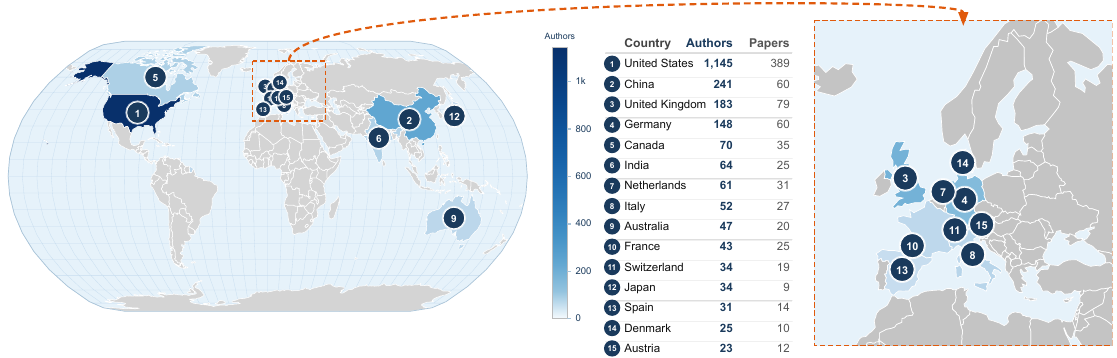}
      \caption{\textbf{(a)}}
      \label{fig:sub1a}
  \end{subfigure}

  \vspace{1em}

    \hspace{-8em}
  \begin{subfigure}[t]{\textwidth}
      \centering
      
      \includegraphics[width=1.2\textwidth]{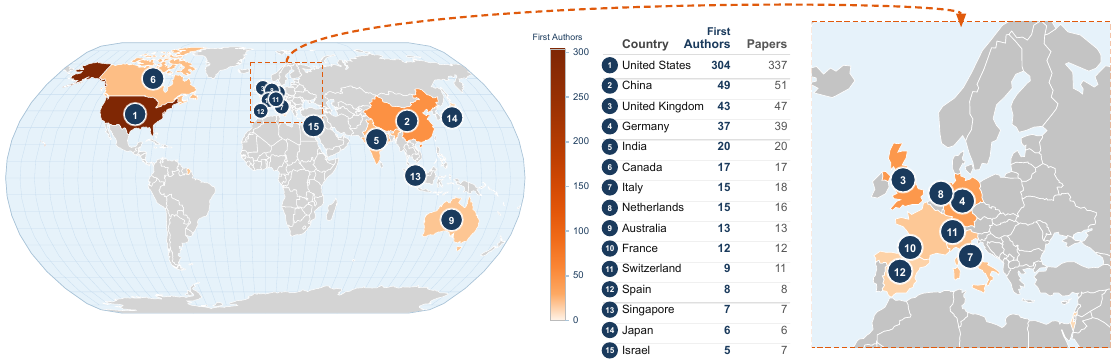}
      \caption{\textbf{(b)}}
      \label{fig:sub1b}
  \end{subfigure}
  \phantomcaption   % increments the figure counter invisibly, no printed caption
\end{figure*}

% Page 2: subfigure (c) + full caption at the end
\begin{figure*}[p]
  \ContinuedFloat
  \centering
  \begin{subfigure}[t]{1\textwidth}
      \centering
      \includegraphics[width=\textwidth]{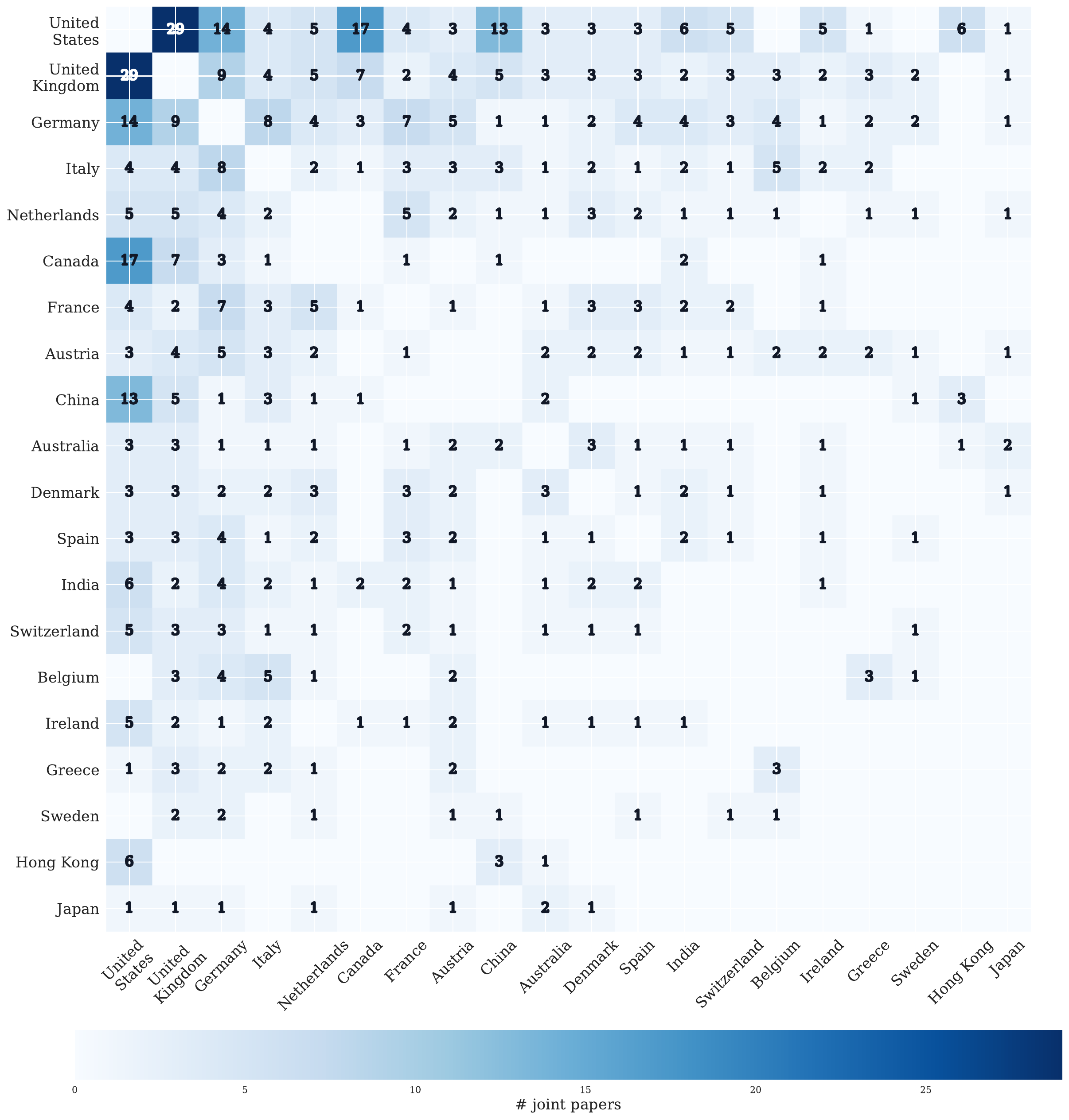}
      \caption{\textbf{(c)}}
      \label{fig:cross_country_collab}
  \end{subfigure}
  \caption{Country-level authorship and publication patterns. \textbf{(a)} Total unique author $\times$ paper counts by country. \textbf{(b)} First-author counts and papers by country. \textbf{(c)} Country-wise collaboration counts.}
  \label{fig:country_authorship}
\end{figure*}

\begin{figure*}[htb] 
\centering                  
\begin{subfigure}[t]{0.8\textwidth}                             
\centering                                                 
\includegraphics[width=\textwidth]{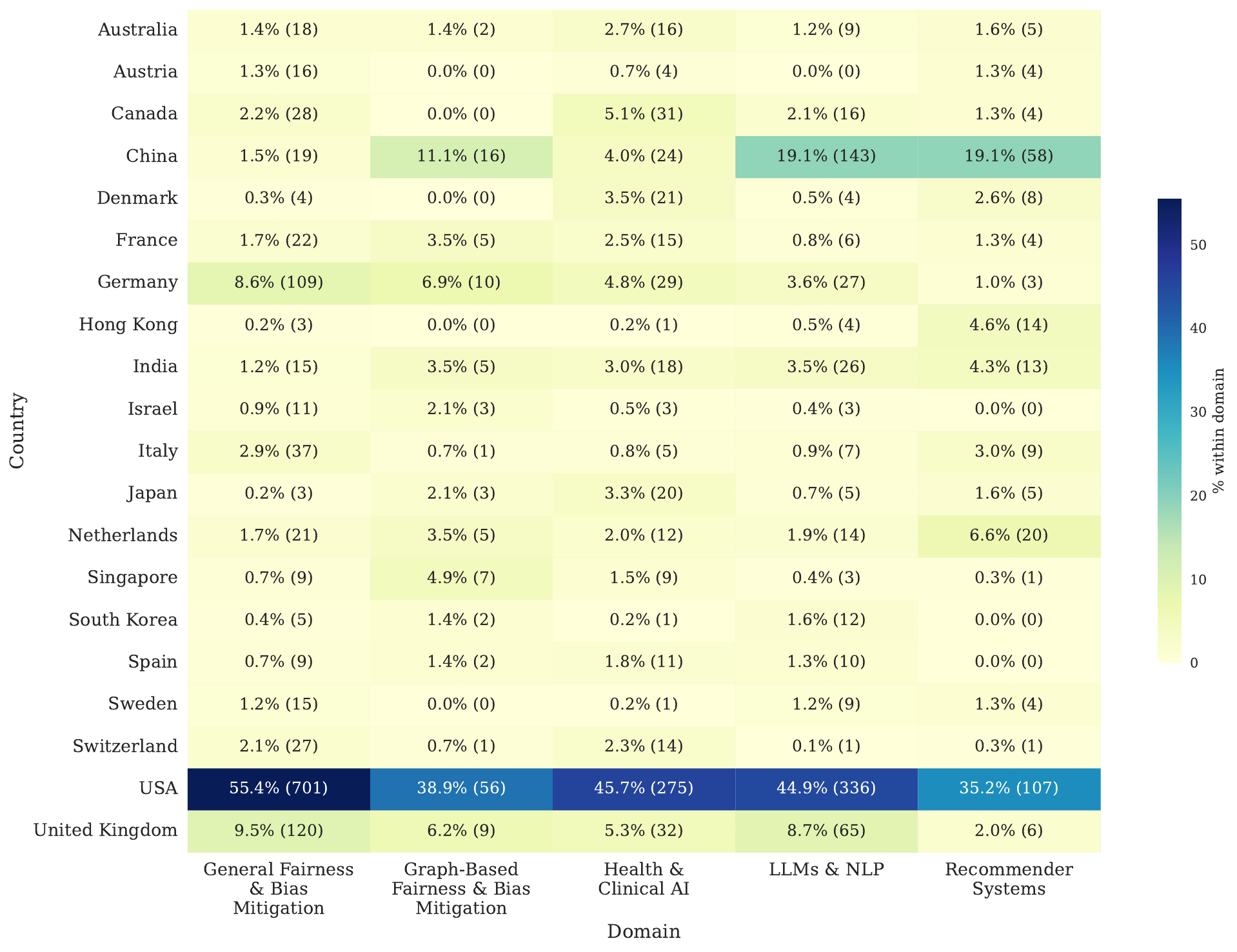}
\caption{\textbf{(a)}}              
\label{fig:country-all}                          
\end{subfigure}                                   
\hfill                                                
\begin{subfigure}[t]{0.8\textwidth}                   
\centering                                                             \includegraphics[width=\textwidth]{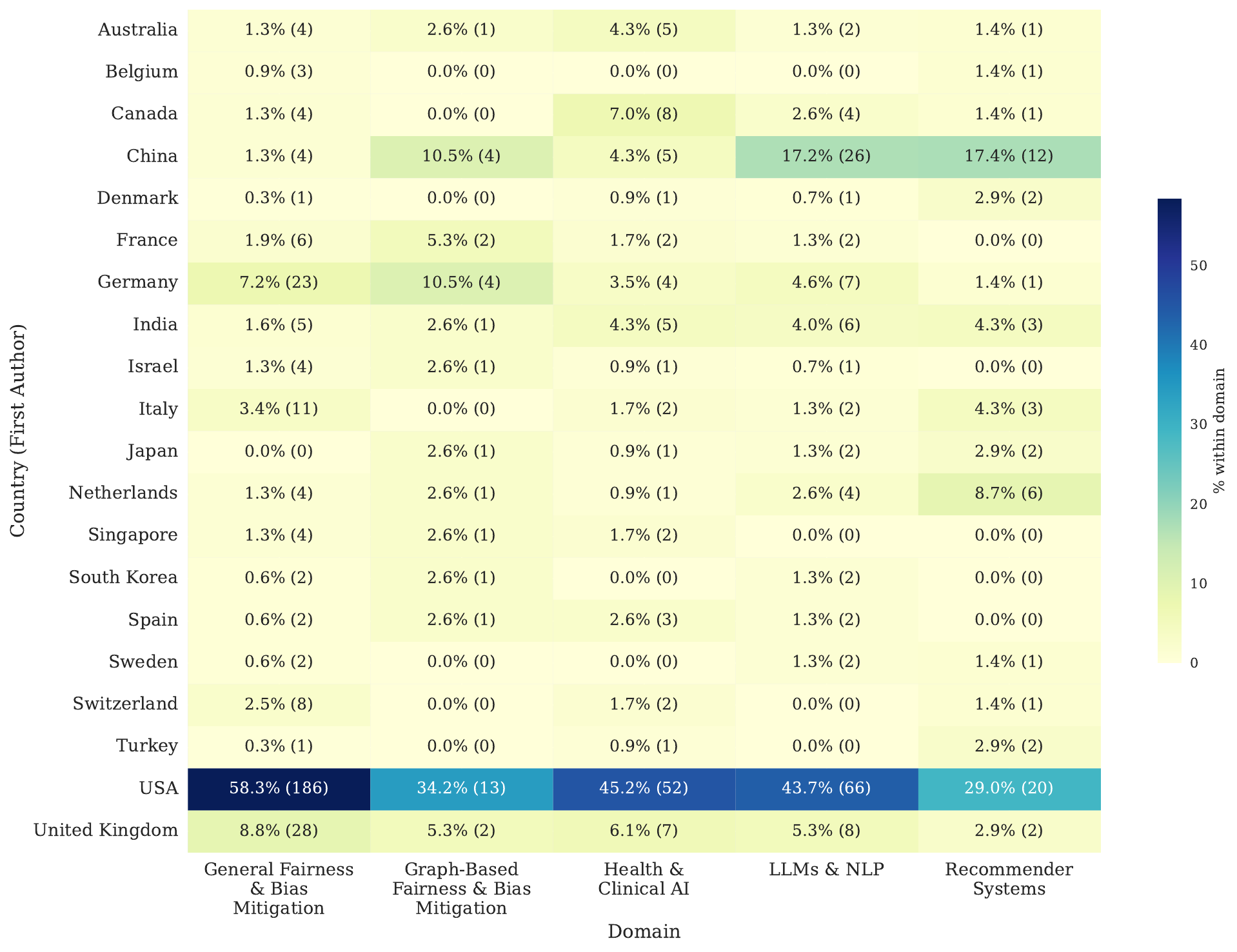}    
\caption{\textbf{(b)}}            
\label{fig:country-first}                      
\end{subfigure} 
\caption{Top 20 country-wise contributions to AI bias research across domains. \textbf{(a)} All-author counts and percentages. \textbf{(b)}        
  First-author counts and percentages. The United States leads across all domains; China's contribution is concentrated in LLMs \& NLP and Recommender Systems.} 
  \label{fig:country}                                                        
\end{figure*}  
  
The United States~(US) dominates AI bias research production, contributing 1{,}145 unique authors across 389 papers in total authorship and 304 first authors across 337 first-authored papers (Fig.~\ref{fig:sub1a},~\ref{fig:sub1b}), indicating not only the largest collaborative base but also the strongest research leadership. China ranks second in total authorship~(241 authors, 60 papers; 49 first authors, 51 papers), followed by the United Kingdom~(UK)~(183 authors, 79 papers; 43 first authors, 47 papers) and Germany~(148 authors, 60 papers; 37 first authors, 39 papers). Notably, the UK surpasses China in paper count~(79 vs.\ 60) despite fewer total authors, indicating smaller average team sizes. The Global South, including South Asia, Africa, and Latin America, remains substantially underrepresented in both authorship metrics. India shows the highest contribution among underrepresented regions~(64 authors, 25 papers; 20 first authors, 20 papers), while Italy~(52 authors, 27 papers; 15 first authors, 18 papers) surpasses Canada and the Netherlands in first-authored output. Both India and Italy remain far below the dominant contributors. Moreover, the Gini coefficient for first-authored papers across the 52 contributing countries is  0.7842 (95\% CI [0.7395, 0.7877] (0 = perfectly equal distribution, 1 = maximal concentration)), with the US alone accounting for 48.7\% of first-authored papers in the corpus, indicating high concentration of research output within a small number of countries.

From a domain-level perspective, the US leads in contribution across all domains in both all-author~(Fig.~\ref{fig:country-all}) and first-author~(Fig.~\ref{fig:country-first}) analyses, whereas China emerges as the second-largest contributor, particularly in LLMs \& NLP (19.1\% all-author, 17.2\% first-author) and Recommender Systems (19.1\%, 17.4\%). Importantly, the US alone accounts for more than half of all output in the foundational domain, i.e., General Fairness \& Bias Mitigation, holding 55.4\% of all-author and 58.3\% of first-author publications. However, Chinese contribution drops to 4.0\% and 1.5\% for all-author contributions in Health \& Clinical AI and General Fairness, respectively, and to 4.3\% and 1.3\% for first-author contributions, suggesting that China's research is oriented toward technical and generative AI rather than foundational fairness frameworks or clinical applications. Health \& Clinical AI is the most internationally balanced domain: in all-author contributions, Canada~(5.1\%), the UK~(5.3\%), Germany~(4.8\%), and China~(4.0\%) each contribute comparable shares behind the US (45.7\%), and first-author leadership similarly diversifies, with Canada~(7.0\%) and the UK~(6.1\%) following the US~(45.2\%). The UK is the second-largest contributor in the General Fairness domain (9.5\% all-author and 8.8\% first-author), followed by Germany (8.6\% and 7.2\%). In the Graph-Based Fairness domain, China and Germany jointly rank second, each accounting for 10.5\% of first-author publications.
Asian participation from countries such as India, Japan, Singapore, and Hong Kong, is also visible across multiple domains, suggesting broader participation beyond the dominant contributors.

A chi-square test of independence confirms that domain distribution varies significantly by country ($\chi^2 = 123.78$, $df = 40$ [11 countries × 5 domains], Cram\'{e}r's $V = 0.231$, a weak-to-moderate effect). Since $61.8\%$ (34 of 55) of country–domain pairs had expected counts below five, the $p$-value was estimated via Monte Carlo permutation ($10{,}000$ permutations, $p < 0.0001$), which does not rely on minimum expected cell counts, rather than the asymptotic approximation which assumes expected cell counts are sufficiently large. This suggests that research priorities in AI bias are not uniform across countries and are significantly skewed toward particular thematic areas.

 The United States exhibits the highest level of international collaboration (Fig.~\ref{fig:cross_country_collab}). The most frequent collaboration is between the US and the UK~(29 joint papers), followed by the US and Canada~(17), the US and Germany~(14), and the US--China~(13). A prominent European collaboration cluster centers around Germany, which exhibits strong collaboration links with UK (9 joint papers), Italy (8), and France (7). The UK further bridges this network with connections to Italy (4 joint papers), Austria (4), and the Netherlands (5). China's collaboration is heavily focused on Western hubs, primarily the US (13) and the UK (5), followed by partnerships with Hong Kong (3) and Italy (3).

The Global South is largely underrepresented within the collaboration network, thereby reinforcing the structural underrepresentation identified in country-level authorship and domain-wise contribution patterns.
The collaboration structure suggests that knowledge exchange in AI bias research occurs predominantly within a small number of Western nations, potentially limiting the integration of non-Western perspectives into the mainstream discourse.

\subsection*{Institutional Contributions}

At the institutional level, a domain-specific specialization in research output is observed. The institutional Gini coefficient is 0.5885 across 703 institutions when measured by all-author affiliations, and 0.3867 across 369 institutions
when restricted to first-author affiliations. This indicates that first-author leadership is more broadly distributed than collaborative participation. However, these aggregate figures mask the domain-level variations.

\begin{figure*}[htb]                                                 \centering                                                          \begin{subfigure}[t]{0.9\textwidth}                                 \centering                                                          \includegraphics[width=\textwidth]{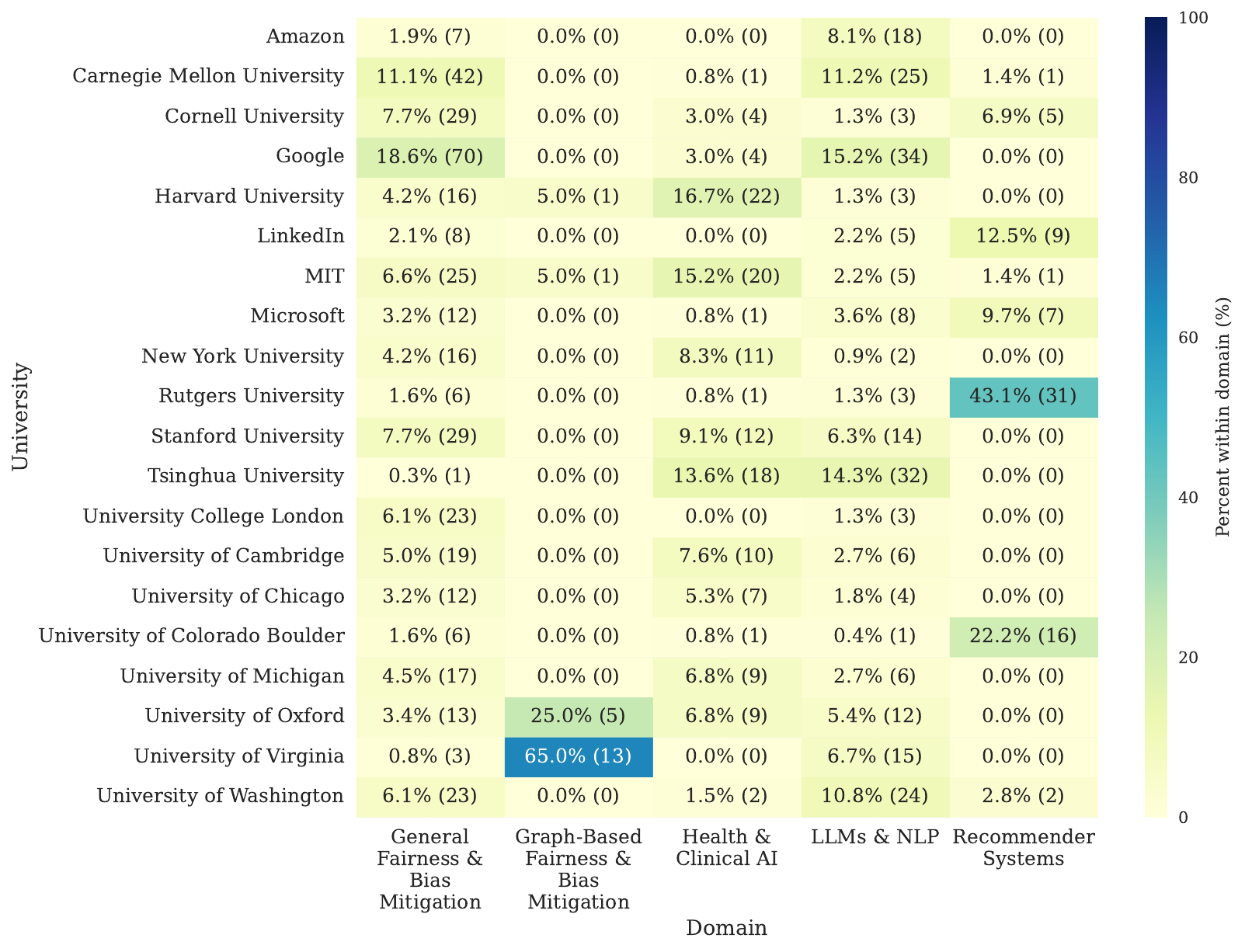}
          \caption{\textbf{(a)}}                                                                                                          \label{fig:institution-all}                                      \end{subfigure}   
          \hfill                                                           \begin{subfigure}[t]{0.9\textwidth}
          \centering                                                      \includegraphics[width=0.9\textwidth]{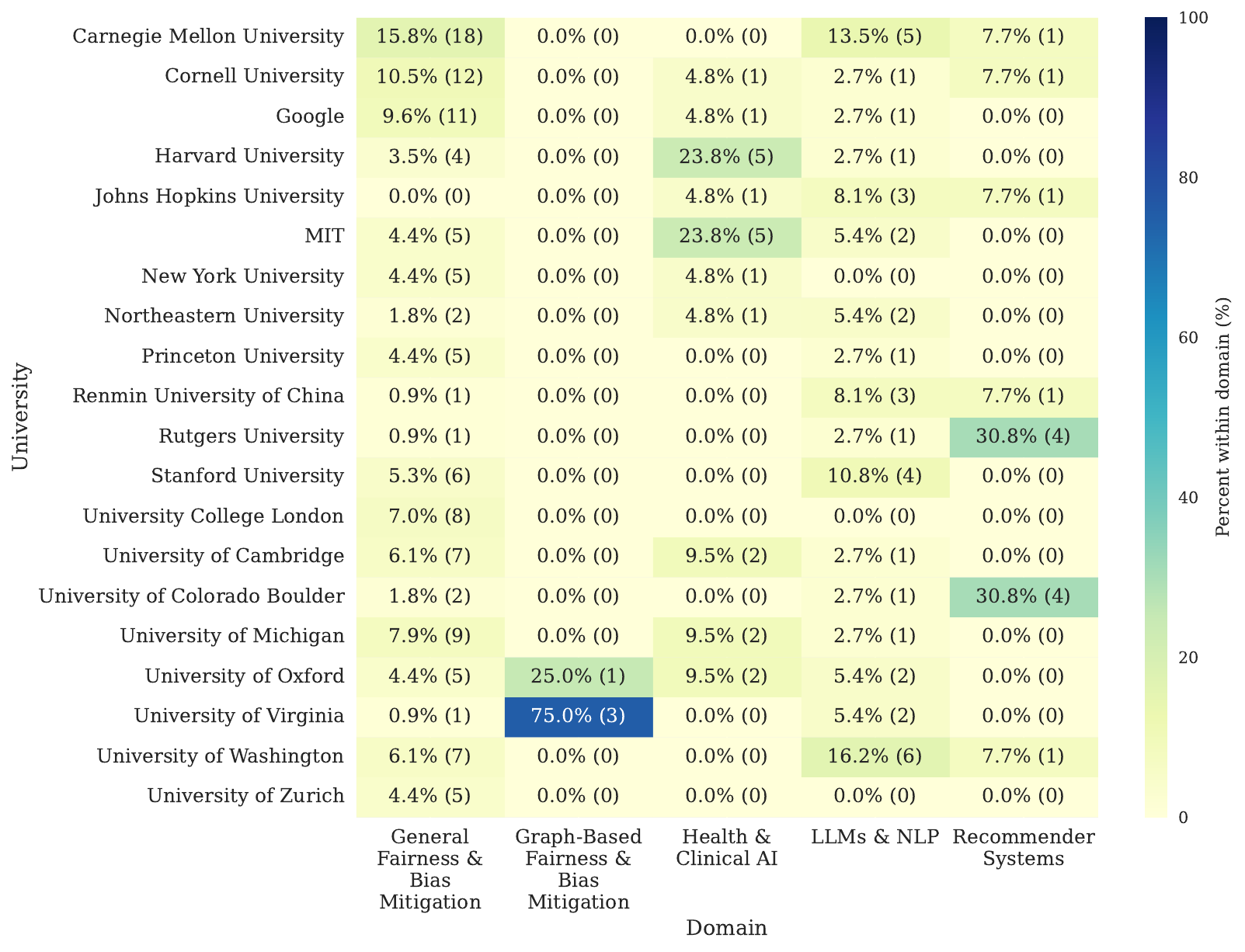}                              \caption{\textbf{(b)}}
          \label{fig:institution-first}                                \end{subfigure}
  \caption{Top 20 institutional contributions to AI bias research across domains. \textbf{(a)} All-author counts. \textbf{(b)} First-author counts.\\ Percentages denote each institution's share among the top-20 institutions shown, per domain.
  } 
  \label{fig:institution}                                             \end{figure*}

In General Fairness \& Bias Mitigation, Google leads all-author contributions at 18.6\%, followed by Carnegie Mellon~(11.1\%) and Stanford~(7.7\%) (Fig.\ref{fig:institution-all}). However, Google's share drops to 9.6\% in first-author contributions, with Carnegie Mellon~(15.8\%) and Cornell~(10.5\%) leading the first-author output (Fig.\ref{fig:institution-first}), consistent with the lower first-author institutional Gini of 0.3867 compared to 0.5885 for all-author affiliations. This suggests that corporate labs participate heavily in collaborations but academic institutions more frequently lead the research agenda.

LLMs \& NLP exhibits a relatively decentralized first-author leadership: the University of Washington leads at 16.2\%, followed by Carnegie Mellon~(13.5\%) and Stanford~(10.8\%), with Renmin University of China~(8.1\%) among the top contributors. In all-author contributions, Google~(15.2\%) and Tsinghua University~(14.3\%) are the top two, indicating strong industry and Chinese institutional engagement in LLM fairness research as collaborators, while  Renmin University of China's presence among first-author contributors reflects Chinese research leadership in this domain.

The corpus's two smallest domains show extreme concentration. In Recommender Systems Rutgers University is the largest all-author contributor (43.1\%), but shares the first-author leadership equally with the University of Colorado Boulder (30.8\% each). In Graph-Based Fairness,
research contribution among the corpus's leading institutions is confined almost entirely to two institutions: the University of Virginia (65.0\% all-author, 75.0\% first-author) and Oxford (25.0\% all-author, 25.0\% first-author). The near complete absence of other leading institutions indicates that bias research in these domains is driven by a small number of research groups, which risks constraining the research agenda to the priorities, methods, and problem framings of those groups. Contribution from Asian countries, including India, Japan, Singapore, and Hong Kong, which was visible at the national level (Fig.~\ref{fig:country}) is not reflected in institutional rankings, indicating that research activity in these countries is distributed rather than concentrated within a small number of institutions.

\subsection*{Citation Analysis and Pattern}

Citation analysis was conducted using 2 lenses: global citation counts sourced from OpenAlex, measuring cross-disciplinary reach,
and internal citations, measuring how many other papers within our 692-paper corpus cite a given work, reflecting the field’s own intellectual dependencies.
The global citation distribution is heavily right-skewed: the corpus-wide  median is 9 and the mean is 93.5, indicating that a small subset of highly cited articles accounts for a disproportionate share of total citations. 

Global citation impact differs significantly across domains
(Kruskal-Wallis $H = 34.55$, $p = 5.75 \times 10^{-7}$). General Fairness \& Bias Mitigation has the highest median global citations at 13 (95\% CI: 9 to 16), followed by Health \& Clinical AI at 10 (95\% CI: 4 to 20), LLMs \& NLP at 6 (95\% CI: 4 to 10), Recommender Systems at 3 (95\% CI: 1 to 7), and Graph-Based Fairness at 2.5 (95\% CI: 0 to 4) (Fig.~\ref{fig:citation-dist-domain}). Post-hoc Dunn's test with Bonferroni correction confirms that General Fairness receives
significantly more citations than LLMs \& NLP ($p = 0.0017$), Recommender Systems ($p = 0.0004$), and Graph-Based Fairness ($p = 0.0010$).
Health \& Clinical AI does not show significant difference from LLMs \& NLP ($p = 0.4806$) or Recommender Systems ($p = 0.0562$), but does differ significantly from Graph-Based Fairness ($p = 0.0372$); its wide confidence interval (95\% CI: 4 to 20 citations) reflects high variability in global citation impact.
The compressed lower quartiles across all domains indicate that the
majority of papers attract few citations, while the upper tails are driven by a handful of widely cited works.

When ranked by cumulative citation count, the internal and global citations follow almost the same domain ordering.
The global ordering is: General Fairness $>$ Health $>$ LLMs $>$ Recommender Systems $>$ Graph,
whereas in the within-corpus ranking, LLMs \& NLP and Health \& Clinical AI are swapped (Fig.~\ref{fig:citation-internal-global}), indicating that LLMs \& NLP papers are more heavily cited within the corpus than their global ranking suggests.
Health \& Clinical AI has the highest global-to-internal ratio (76.7:1), indicating the broadest cross-disciplinary reach, followed by LLMs \& NLP
(50.1:1). Recommender Systems has the lowest ratio (23.1:1), followed by General Fairness \& Bias Mitigation (28.1:1) and Graph-Based Fairness (35.8:1).
For Recommender Systems and Graph-Based Fairness, low ratios reflect limited external reach, suggesting comparatively self-contained research, as for General Fairness, the low ratio indicates the field's heavy internal reliance on its foundational work.
At the country level, publication volume is not significantly correlated with citation impact (Fig.~\ref{fig:citation-regional}).
The US, despite leading in total output (389 papers), has a median citation count of 13, behind Canada (21), Austria (19), Spain (17), and the UK (14)  suggesting that output from some smaller contributors is more selectively influential. The UK, which ranks second in total output, achieves a higher median citation count (14) than the US despite contributing far fewer papers (389 vs 70 papers), indicating a stronger per-paper research impact.
China and India, which rank among the top 10 by volume, show the 2 lowest median citations, suggesting high output but lower per-paper citation impact.

A Spearman rank-order correlation between country-level publication volume and median citation count yielded no significant association~($\rho =0.073$, $p = 0.831$, $n = 11$ countries with $\geq 10$ papers), confirming that research volume and citation impact are decoupled at the country level.
           
The 15 most impactful papers, with the highest citation count, belong to three major domains: General Fairness \& Bias Mitigation (8 papers), Health \& Clinical AI~(4 papers), and LLMs \& NLP~(3 papers)~(Fig.~\ref{fig:citation-top15}), consistent with their leading position in overall citation volume (Fig.~\ref{fig:citation-internal-global}).
The top three most-cited papers in the corpus, Obermeyer et al.~\cite{obermeyer_dissecting_2019}~(5,913), Bender et
al.~\cite{bender_dangers_2021}~(5,297), and Dwork et al.~\cite{dwork_fairness_2012}~(3,331), lead the remainder by a wide margin, with the highest-cited paper exceeding
the fifteenth-ranked entry by nearly six-fold.
Graph-Based Fairness and Recommender Systems are entirely absent, indicating that high-impact work in the field remains concentrated within the domains that also lead in overall citation volume~(Fig.~\ref{fig:citation-internal-global}).

% Page 1: subfigures (a) and (b) — no main caption
\begin{figure*}[p]
  \centering
  \begin{subfigure}[t]{1\textwidth}
      \centering
      \includegraphics[width=1\textwidth, keepaspectratio]{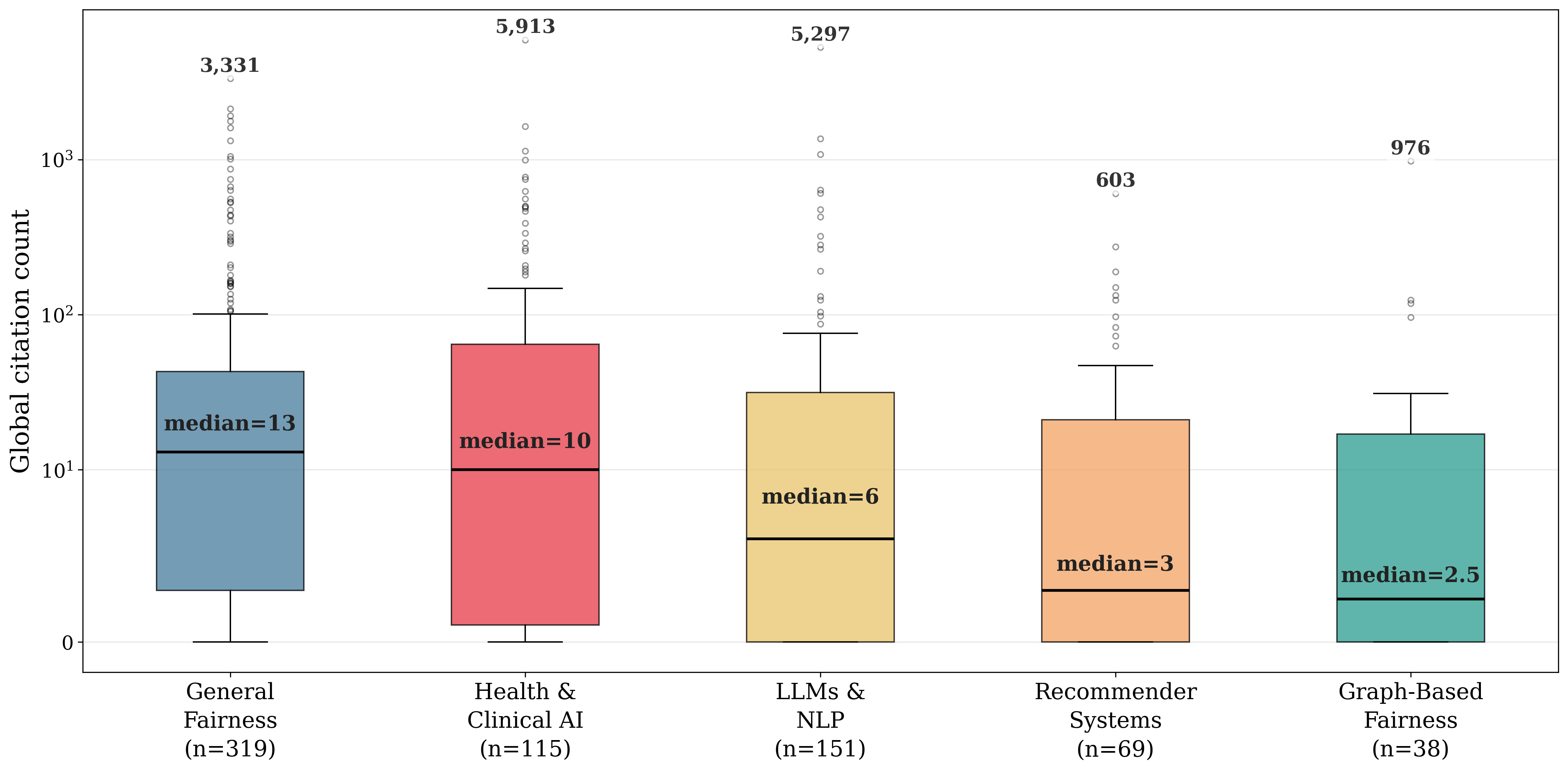}
      \caption{\textbf{(a)}}
      \label{fig:citation-dist-domain}
  \end{subfigure}

  \vspace{1em}

  \begin{subfigure}[t]{1\textwidth}
      \centering
      \includegraphics[width=1\textwidth, keepaspectratio]{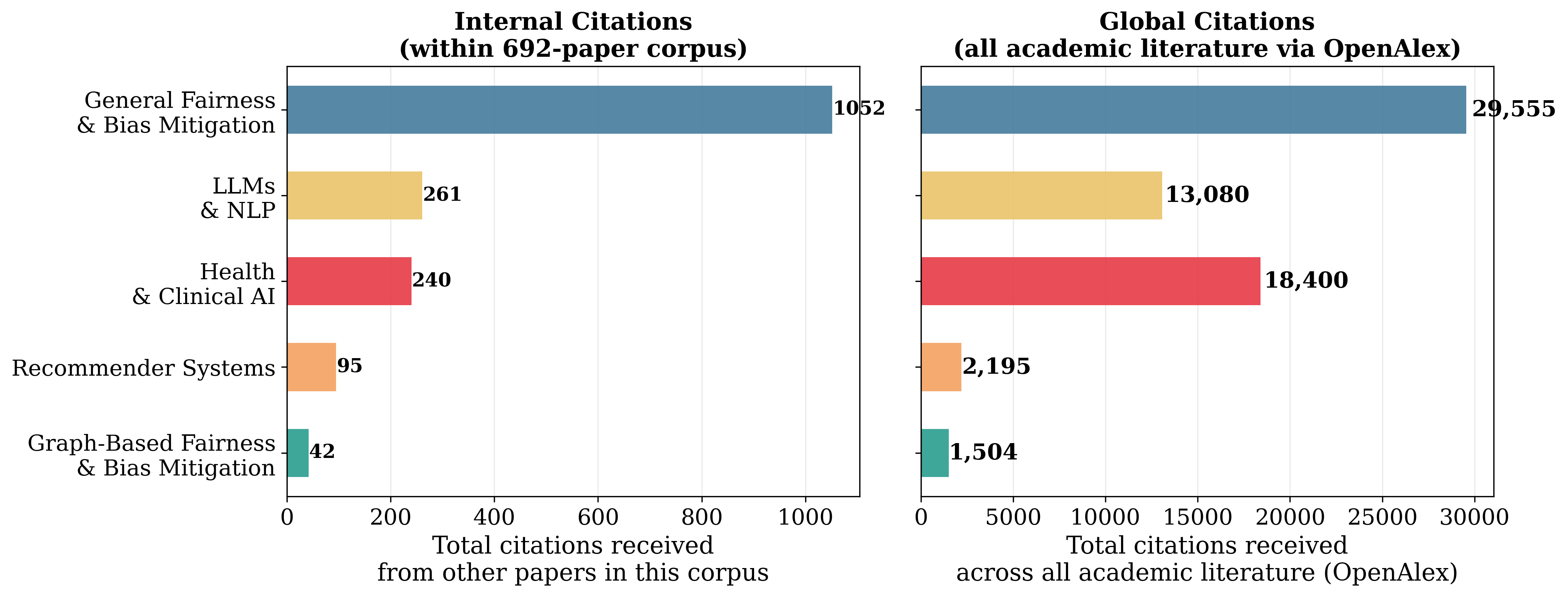}
      \caption{\textbf{(b)}}
      \label{fig:citation-internal-global}
  \end{subfigure}
  \phantomcaption
\end{figure*}

% Page 2: subfigures (c) and (d) + full caption at the end
\begin{figure*}[p]
  \ContinuedFloat
  \centering
  \begin{subfigure}[t]{1\textwidth}
      \centering
      \includegraphics[width=1\textwidth]{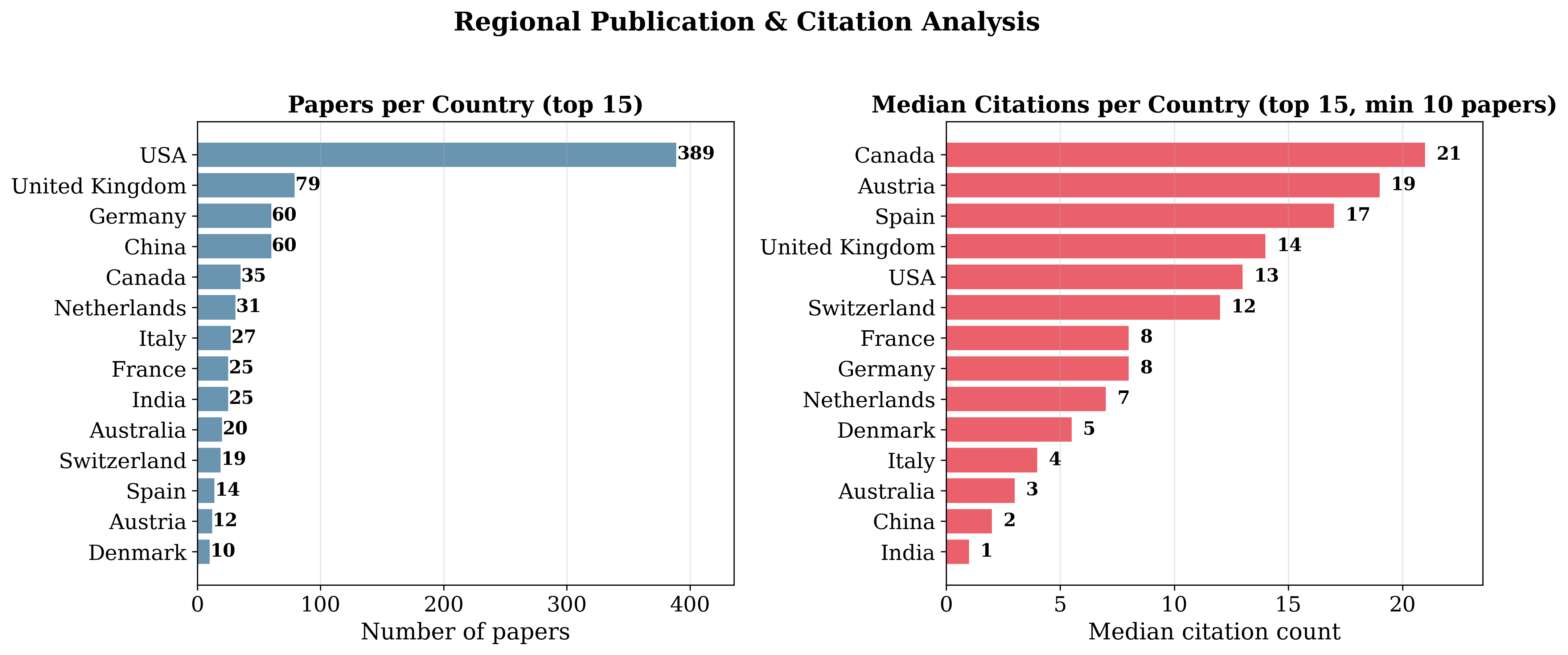}
      \caption{\textbf{(c)}}
      \label{fig:citation-regional}
  \end{subfigure}

  \vspace{1em}

  \begin{subfigure}[t]{1\textwidth}
      \centering
      \includegraphics[width=1\textwidth]{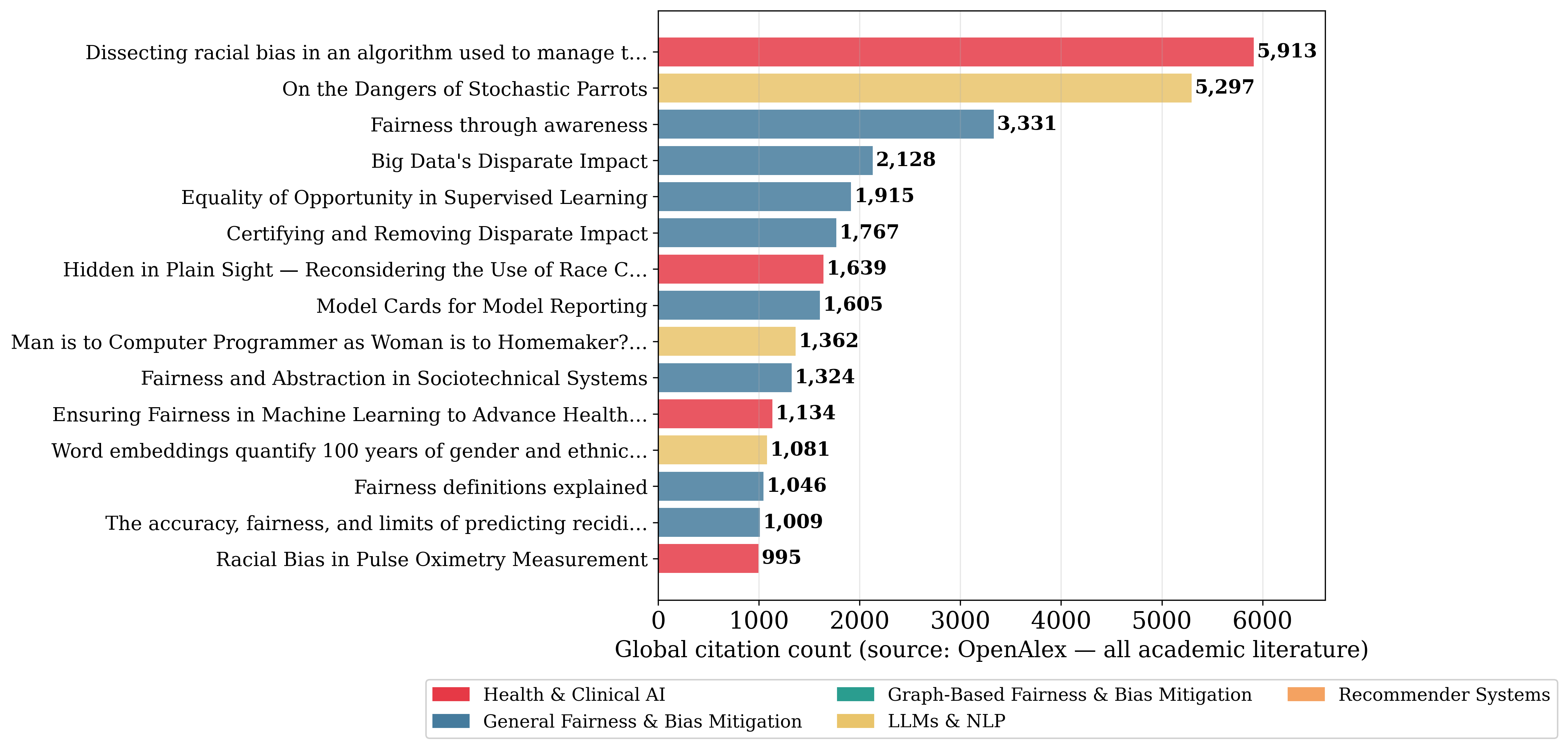}
      \caption{\textbf{(d)}}
      \label{fig:citation-top15}
  \end{subfigure}
  \caption{Citation analysis. \textbf{(a)} Global citation count distribution across domains (symlog scale). \textbf{(b)} Total internal
  (within-corpus) and global (OpenAlex) citation volume per domain. \textbf{(c)} Countries ranked by paper count and median global citations
  (countries with $\geq$10 papers). \textbf{(d)} Top 15 most-cited papers by global citation count (OpenAlex).}
  \label{fig:citations}
\end{figure*}

\subsection*{Semantic landscape}

\begin{figure*}[p]
  \centering

  \begin{subfigure}[t]{0.48\textwidth}
    \centering
    \includegraphics[
      width=\textwidth,
      height=0.30\textheight,
      keepaspectratio
    ]{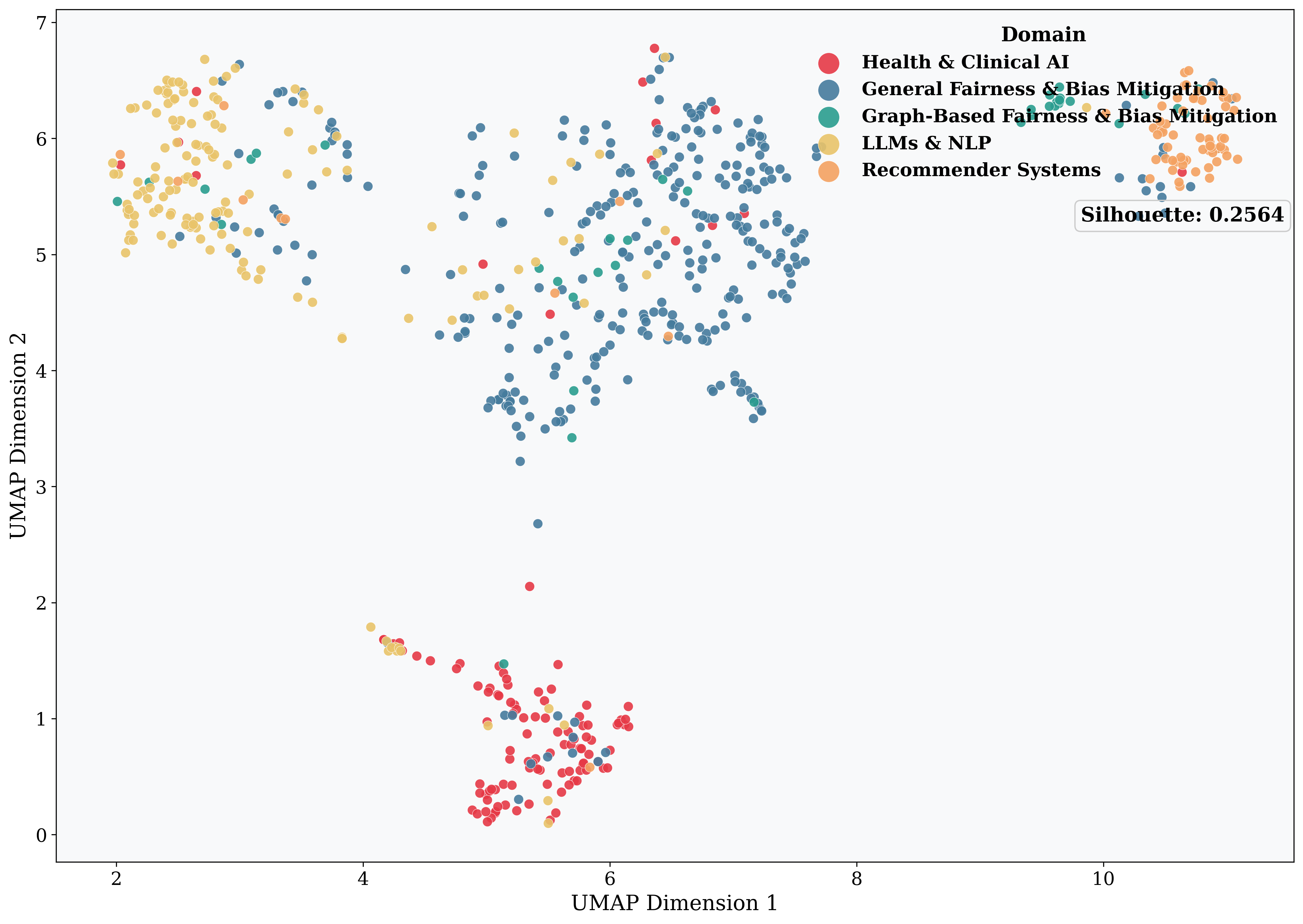}
    \caption{\textbf{(a)}}
    \label{fig:semantic-landscape}
  \end{subfigure}
  \hfill
  \begin{subfigure}[t]{0.48\textwidth}
    \centering
    \includegraphics[
      width=\textwidth,
      height=0.30\textheight,
      keepaspectratio
    ]{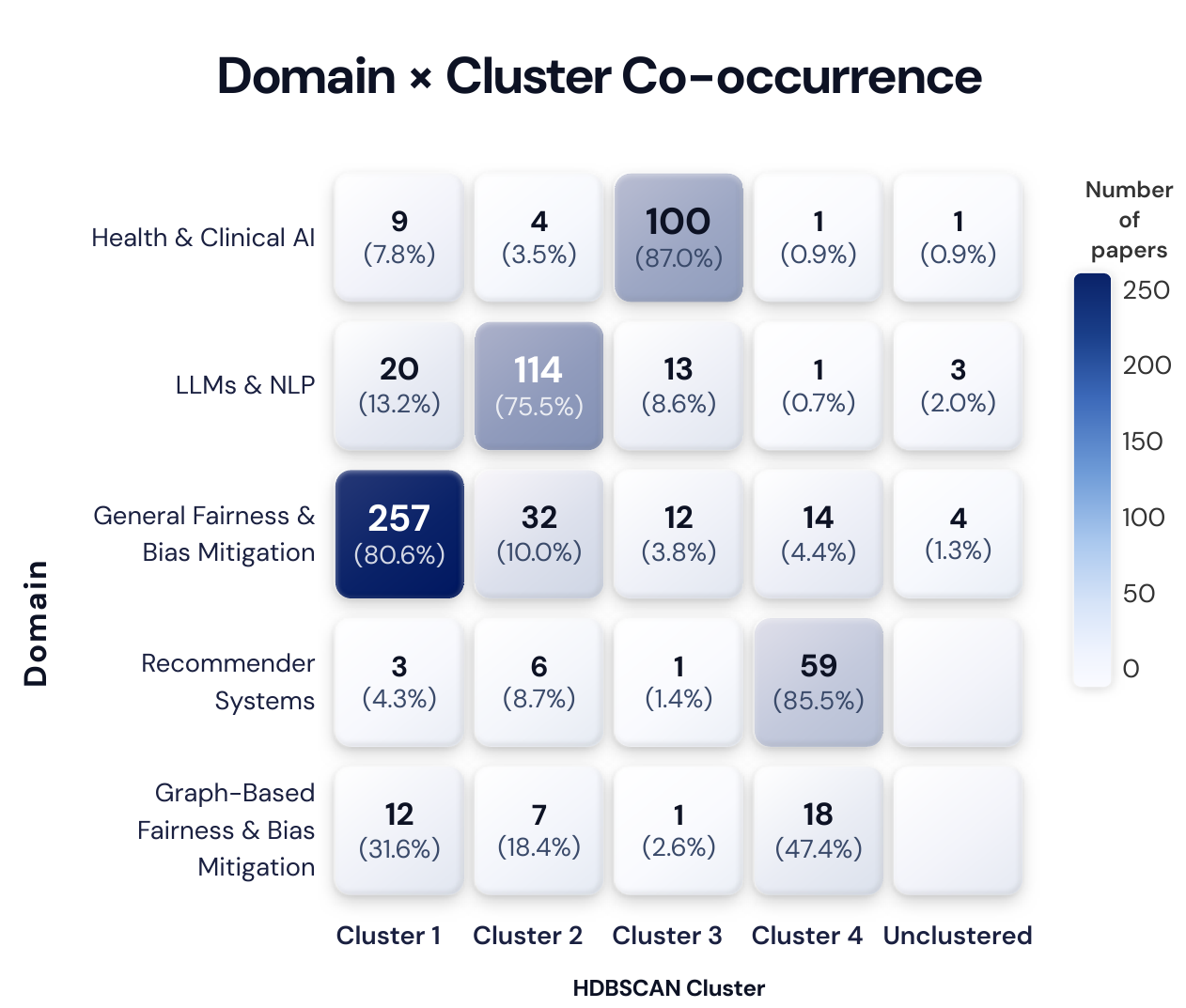}
    \caption{\textbf{(b)}}
    \label{fig:domain-cluster}
  \end{subfigure}

  \vspace{0.75em}

  \begin{subfigure}[t]{0.65\textwidth}
    \centering
    \includegraphics[
      width=\textwidth,
      height=0.32\textheight,
      keepaspectratio
    ]{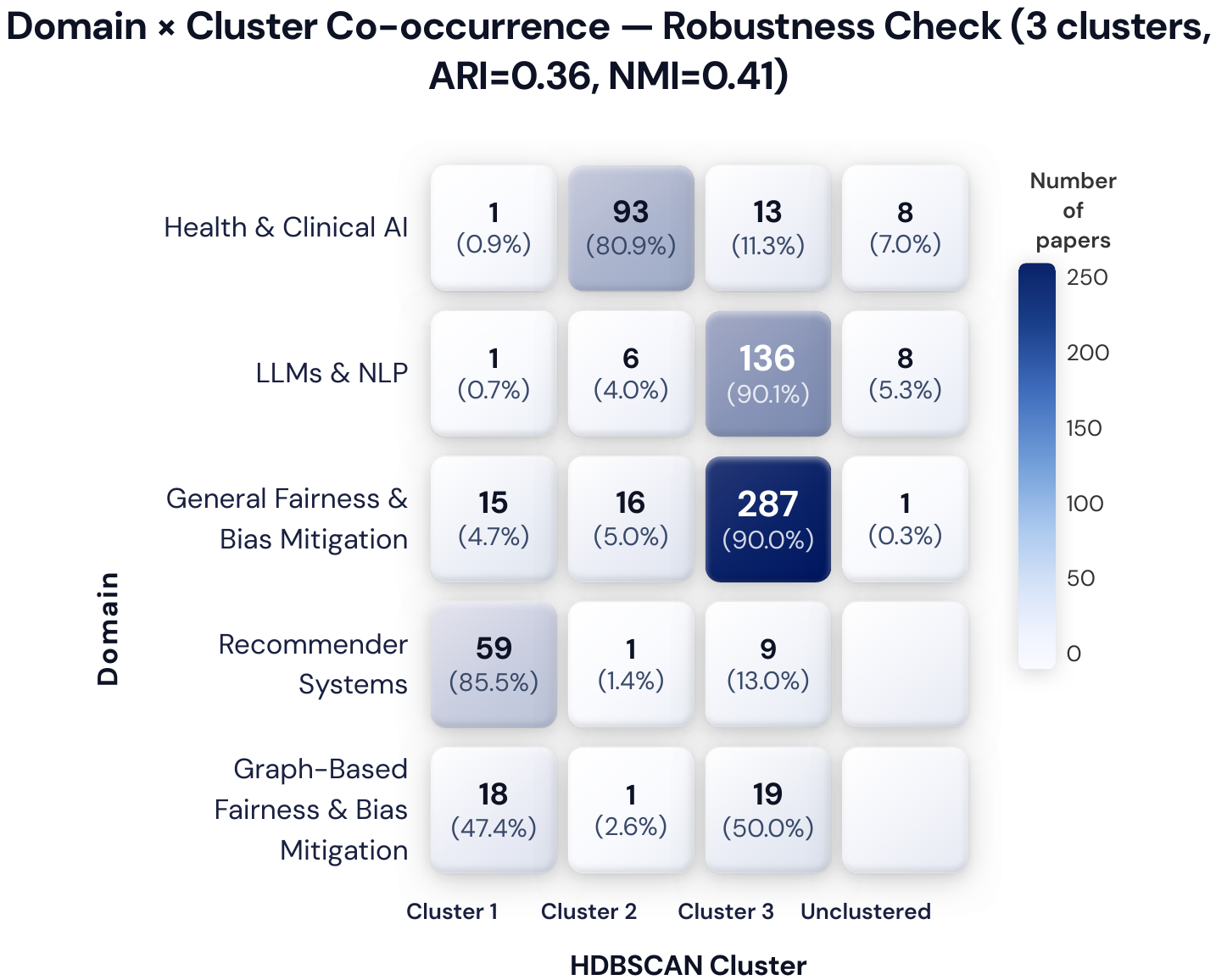}
    \caption{\textbf{(c)}}
    \label{fig:robustness}
  \end{subfigure}

  \caption{Semantic clustering analysis. \textbf{(a)} UMAP projection of
  Sentence-BERT (\texttt{all-mpnet-base-v2}) abstract embeddings
  ($n = 692$), coloured by assigned domain. Silhouette score: 0.2564.
  \textbf{(b)} Domain--cluster co-occurrence matrix using optimized UMAP
  ($n_{\text{neighbors}} = 30$, $\min_{\text{dist}} = 0.0$) and HDBSCAN
  ($\min_{\text{cluster\_size}} = 20$, $\min_{\text{samples}} = 10$)
  parameters. Cells show count and row-normalised percentage.
  \textbf{(c)} Robustness check: domain--cluster co-occurrence under
  default UMAP parameters ($n_{\text{neighbors}} = 15$,
  $\min_{\text{dist}} = 0.1$) with HDBSCAN
  ($\min_{\text{cluster\_size}} = 20$, $\min_{\text{samples}} = 10$);
  ARI = 0.36, NMI = 0.41.}
  \label{fig:umap}
\end{figure*}

UMAP~(Uniform Manifold Approximation and Projection) projection of all 692 paper abstracts was performed using Sentence-BERT embeddings. Three embedding models were evaluated via grid search over UMAP hyperparameters. MPNet achieved the best silhouette score (0.2564), outperforming  MiniLM~(0.2137) and SPECTER~(0.2052), and was selected as the default projection. The resulting embedding's scatterplot is shown in~Fig.~\ref{fig:semantic-landscape}.

HDBSCAN clustering on the MPNet projection~(UMAP: $n_\text{neighbors} = 30$, $\min_\text{dist} = 0.0$; HDBSCAN: $\min_\text{cluster\_size} = 20$, $\min_\text{samples} = 10$) produced four semantically coherent clusters with eight noise points~(1.2\%) and a clustering silhouette score of 0.6852~(ARI~(Adjusted Rand Index) $= 0.53$, NMI~(Normalized Mutual Information)~$= 0.47$, substantially above chance).
While the embedding silhouette (0.2564) is low in absolute terms, indicating partial rather than complete domain separation in the projection, it was used solely as a criterion for embedding model selection. The clustering silhouette, along with ARI and NMI scores, confirms that the resulting clusters are internally coherent. Some overlap is expected given the thematic breadth of AI bias research.

The resulting clusters showed substantial alignment with the five manually-assigned domains, demonstrating the semantic coherence of our domain classification~(Fig.~\ref{fig:domain-cluster}). Health \& Clinical AI and Recommender Systems each map predominantly to a single cluster~(87.0\% and~85.5\% overlap, respectively). LLMs \& NLP concentrates~88.7\% of papers~(134 of 151) in two semantically adjacent clusters.
General Fairness \& Bias Mitigation is a domain with meaningful representation (ranging from 3.8\% to 80.6\%) across all four clusters, consistent with its cross-cutting scope.
Graph-Based Fairness is the most scattered, split across all four clusters (47.4\%, 31.6\%, 18.4\%, 2.6\%) with no single cluster capturing a majority - unlike the other domains, suggesting that it functions as a methodology applied across application domains rather than a self-contained research area.

To confirm that this alignment is not an artifact of UMAP hyperparameter tuning, we re-ran the analysis using default UMAP parameters ($n_\text{neighbors} = 15$, $\min_\text{dist} = 0.1$)\footnote{\url{https://umap-learn.readthedocs.io/en/latest/parameters.html}} with the same HDBSCAN configuration~($\min_\text{cluster\_size} = 20$, $\min_\text{samples} = 10$), without domain-label-based selection (Fig.~\ref{fig:robustness}). This produced three clusters, with comparable domain overlap: 
Recommender Systems~(85.5\%), Health \& Clinical
AI~(80.9\%), LLMs \& NLP~(90.1\%), and General Fairness \& Bias Mitigation~(90.0\%) each mapped predominantly to a single cluster~(ARI $= 0.36$, NMI $= 0.41$;  both above chance), confirming that our domain categorization holds independently of projection optimization.

% ============================================================
% DISCUSSION (no subheadings per Nature MI format)
% ============================================================
\section{Discussion}
Our findings suggest that regional concentration in AI-bias research is not only a question of representation but also one of epistemic influence. Fairness is shaped by cultural, social, institutional, and regulatory contexts; therefore, geographically narrow participation can affect which harms are recognized, which populations are prioritized, and which forms of evidence are treated as valid. This is particularly significant in the field's methodological foundations, where fairness definitions, benchmarks, datasets, and mitigation methods are developed and subsequently adapted across application domains. When methodological foundations are produced within a concentrated set of countries and institutions, the assumptions embedded in fairness definitions, benchmarks, and mitigation methods may travel widely without generalizing equally across social and cultural contexts. Our findings show that AI-bias research is geographically concentrated, with the strongest concentration occurring in the methodological layer that informs downstream applications. 

The General Fairness \& Bias Mitigation domain occupies this layer. It is the largest domain in the corpus~(46.1\% of publications), the most cited, semantically cross-cutting, and it is where US dominance is strongest, with US-based authors leading 58.3\% of first-authored output. In first-author counts, this output is driven largely by universities, while all-author participation also reflects the prominent role of companies such as Google. Because this domain supplies the fairness definitions, benchmarks, and mitigation methods that applied areas inherit, concentration within it has field-wide consequences: inequality in an application area shapes that area, but inequality in the foundational layer shapes the evaluation standards and technical interventions used across the whole field. Methods developed within a narrow set of regions and institutions therefore risk encoding local assumptions as general principles, and strong corporate participation may narrow the agenda further, privileging the problems, datasets, and deployment contexts aligned with commercial platforms. The structure of participation is not uniform across domains. Health \& Clinical AI is the most internationally balanced domain, with Canada, the UK, Germany, and China each contributing comparable shares behind the US, suggesting that clinical fairness is recognized as a globally shared concern. Other domains show more distinctive national profiles.  China's contribution concentrates on LLMs \& NLP and Recommender Systems, reflecting a deployment-oriented research profile with comparatively little output in foundational fairness theory. A chi-square test confirms that domain distribution varies significantly by country (Cram\'er's $V = 0.23$, a weak-to-moderate association), indicating that national contributions to AI-bias research are therefore not evenly distributed across thematic areas, but reflect different research priorities and institutional strengths. These patterns highlight that AI-bias research is not a monolithic field. Its inequalities are domain-dependent, and the areas that define the field's shared methodological vocabulary remain among the most geographically concentrated. 

Moreover, the collaboration structure appears to reinforce, rather than diffuse, existing concentration. Knowledge exchange is concentrated within a Western cluster centered on the US, the UK, Canada, and Germany. China's collaborations are also directed largely toward these same hubs, while countries in the Global South are sparsely represented in the co-authorship network. India is a notable case: despite its large higher-education system, technical workforce, and growing AI research capacity, it remains substantially underrepresented in the collaboration structure observed here. This pattern is consistent with cumulative-advantage and preferential-attachment dynamics associated with the Matthew effect~\cite{10.1098/rsif.2014.0378, doi:10.1126/science.159.3810.56}. Collaboration tends to flow toward already central actors, thereby reinforcing established centres of activity and limiting the incorporation of perspectives from less central regions into the mainstream AI-bias research discourse.

Publication volume and citation influence are not equivalent in our analysis. We find no significant association between a country's publication output and its median citation impact~($\rho = 0.073$, $p = 0.831$, $n = 11$ countries with $\geq 10$ papers). Several countries with smaller publication volumes achieve higher median citation counts than the dominant producers, indicating that quantitative dominance does not necessarily translate into greater per-paper influence. This result complicates a simple interpretation of output as influence. Countries that produce fewer papers may still contribute work that is highly visible within the field, whereas high-volume producers may exert influence through breadth, agenda-setting capacity, and institutional centrality rather than citation impact alone. However, this decoupling should be interpreted cautiously. Median citation estimates for smaller contributors are based on relatively few papers and are therefore less stable. Citation-based influence also captures only one dimension of impact and may not fully reflect methodological uptake, benchmark adoption, or downstream use across domains.

The semantic analysis corroborates the manual taxonomy: the resulting clusters show substantial alignment with the five manually assigned domains, indicating that these labels capture actual thematic domains rather than imposed categories. Health \& Clinical AI and Recommender Systems each map cleanly onto a single cluster, suggesting that these domains are semantically well defined. Graph-Based Fairness \& Bias Mitigation, on the other hand, provides a contrast. It does not form a distinct cluster of its own, but instead appears across nearly all other clusters. This pattern suggests that graph-based fairness functions less as a separate application domain than as a methodological lens applied wherever relational structure is relevant. Its semantic dispersion is therefore consistent with both its smaller share of the corpus and its concentration within a limited number of research groups.

Prior works have explored specific domains such as
AI life science~\cite{Schmallenbach2024}, AI based healthcare~\cite{Alberto2024}, and Psychology~\cite{SanchesdeOliveira2023,Henrich2010}, demonstrating how geographic and institutional concentration in research production
is associated with reduced generalizability of findings.
Our results suggest a similar dynamic in AI bias research: the frameworks, metrics, and benchmarks that define what counts as``fair'' are produced by a narrow set of regions and institutions, primarily in the US and Western Europe, while the populations outside the dominant research geographies remain either absent or minimally present in the research community. This pattern contributes to a persistent gap between formal advances in fairness research and context-specific bias harms observed in deployed AI systems. It is important to note that this conclusion must be interpreted in light of our sampling frame. The corpus is drawn from English-language and Global-North-indexed databases, so the observed underrepresentation partly reflects database coverage and cannot be cleanly separated from a true production gap. Broader coverage may alter the magnitude of the imbalance but is unlikely to reverse its direction. Our data do not directly establish that geographic concentration reduces the global generalizability of mitigation methods. However, our findings in the Nepali cultural context~\cite{pandey2026dualmetricevaluationsocialbias}, where LLM bias appears in forms not fully captured by evaluation frameworks developed elsewhere, illustrate the risk of generalizing locally produced standards across contexts. Addressing AI bias requires broadening not just the methods but the communities that produce them. 

These findings have implications for how fairness methods are developed and validated. Benchmarks, definitions, and mitigation baselines produced within a narrow set of regions and institutions should not be treated as context-neutral by default. The field would benefit from clearer reporting of validation populations and settings, stronger external validation in under-represented sociocultural contexts, and greater collaboration and reviewing capacity in regions currently weakly represented in AI-bias research.

To longitudinally track the evolution of AI bias research, we provide a dynamic interactive atlas~(\url{https://biasatlas.cair-nepal.org}) where the community can explore the corpus, submit new papers for inclusion, and monitor shifts in thematic focus, geographic participation, and citation patterns as the field develops. This allows the structural properties of AI bias research — who produces it, where, in which domains, and with what impact — to be tracked continuously rather than retrospectively.

\section{Limitations}\label{sec:limitations}
We acknowledge several limitations that should be considered when interpreting these findings. First, our corpus is drawn primarily from English-language and Global North-indexed sources, including IEEE Xplore, the ACM Digital Library, Scopus, ScienceDirect, and Engineering Village, supplemented with FAccT proceedings and citation snowballing. This sampling strategy likely under-represents research published in other languages, regional venues, and outlets not covered by major indexing services. As a result, the geographic disparities observed here cannot be fully disentangled from database coverage. Although broader coverage is unlikely to reverse the overall pattern of imbalance, the estimated magnitude should be interpreted as reflecting participation within well-indexed segments of the field rather than the field as a whole. Second, thematic domains were assigned through manual screening. Although semantic clustering supports these assignments, with HDBSCAN showing substantial correspondence with the manually defined domains, the modest embedding silhouette indicates considerable overlap between domains. The categories should therefore be interpreted as analytically useful thematic groupings rather than mutually exclusive. Finally, bibliometric indicators capture recorded scholarly output and citation dynamics, but not the broader sociotechnical processes through which fairness norms are negotiated, adopted, and contested. Our measures should therefore be interpreted as a characterization of the field’s documented structure, rather than a complete account of how fairness knowledge is produced and translated into practice.
% ============================================================
% METHODS (online Methods per Nature MI format)
% ============================================================
\section{Methods}
\label{sec:Methods}
\subsection*{Corpus construction}

We began by formulating a set of search strings centered on key concepts such as bias, artificial intelligence, and decision-making including:  ``Bias" AND ( ``Decision-making" OR  ``Decision making") AND ( ``Artificial Intelligence" OR  ``AI" OR  ``Intelligent") AND ``Systems", with variations targeting title and abstract fields across major digital databases such as IEEE Xplore, ACM Digital Library, Scopus, ScienceDirect, and Engineering Village. Citation snowballing was conducted using the Connected Papers and Litmaps platforms to complement and extend the keyword-based
literature search. The ACM Conference on Fairness, Accountability, and Transparency~(FAccT) proceedings were additionally searched, given the venue's direct relevance to AI bias research. Inclusion criteria required publication from 2015 onward, explicit focus on AI bias, and either peer-reviewed status or, for preprints, confirmed relevance through manual screening.
We limited our search to publications from 2015 onward, covering eleven years of research on contemporary AI systems and methods, except for one seminal paper from 2012 \cite{dwork_fairness_2012}, retained for its impact.
A subset of 69 papers (10.0\%) were accessed via arXiv records, of which 10 are FAccT conference papers confirmed as peer-reviewed. Manual review confirmed topical alignment with research objectives, yielding \textbf{692 papers} classified into five thematic domains: (1)~\textit{Health \& Clinical AI}; (2)~\textit{Recommender Systems}; (3)~\textit{LLMs \& NLP}; (4)~\textit{General Fairness \& Bias Mitigation}; and (5)~\textit{Graph-Based Fairness \& Bias Mitigation}.
Abstracts were retrieved via the OpenAlex API and, when unavailable, were collected manually.

\subsection*{Bibliographic metadata extraction}

Metadata was retrieved via the OpenAlex API, including DOIs, author names, affiliations, countries, citation counts, and open access status. Where records were unavailable or malformed, metadata was supplemented through manual verification.

\subsection*{Author and institution counting}
Both all author and first author counts report unique individuals per country, identified by name matching. ORCID-based deduplication was not applied, so authors with identical name strings are treated as the same individual. First-author counts map to the first-listed author per paper. Country-by-domain heatmaps count unique papers per country-domain cell. Institution-by-domain heatmaps count each author's institutional affiliation independently across all authors in the corpus.

\subsection*{Semantic clustering}
\label{sec:methods_cluster}

All 692 papers had retrievable abstracts which we encoded using Sentence-BERT~(embedding dimensions: 768 for MPNet and SPECTER, 384 for MiniLM).
We used the abstracts as they provide a concise, structured representation of each paper's core contributions, methodology, and findings, sufficient to capture the semantic gist and key terminology needed for thematic clustering.
Three models were evaluated: \texttt{all-mpnet-base-v2}, \texttt{all-MiniLM-L6-v2}, and \texttt{allenai-specter}. Dimensionality reduction was performed using UMAP with a cosine metric and we selected hyperparameters via grid search over $n_\text{neighbors} \in \{5, 10, 15, 30, 50\}$ and $\min_\text{dist} \in \{0.0, 0.05, 0.1, 0.2\}$ (20 combinations per model), using silhouette score as the selection criterion.

UMAP was chosen because its two dimensional projections can preserve both the fine-grained similarity between individual papers and the broader separation between topic groups,
both of which matter since HDBSCAN clusters directly on the projected coordinates, not the original 768-dimensional MPNet embeddings.

Each model was evaluated at its own optimal configuration: MPNet ($n_\text{neighbors} = 30$, $\min_\text{dist} = 0.0$), MiniLM ($n_\text{neighbors} = 15$, $\min_\text{dist} = 0.0$), and SPECTER ($n_\text{neighbors} = 10$, $\min_\text{dist} = 0.05$). MPNet achieved the highest silhouette score (0.2564), outperforming MiniLM (0.2137) and SPECTER (0.2052), and was selected as the default projection.
Clustering was performed on the two-dimensional UMAP projection rather than the original 768-dimensional embeddings, as clustering algorithms rely on meaningful distances between points — a property that breaks down in very high-dimensional spaces where all points tend to become equidistant, also known as the curse of dimensionality.
We clustered the two-dimensional MPNet UMAP projection with HDBSCAN. We preferred HDBSCAN over centroid-based methods such as $k$-means because it does not require specifying the number of clusters, accommodates varying cluster density and shape, and labels outliers as noise rather than forcing them into clusters. This suits our analysis, where the number of semantically coherent groups should emerge from the data rather than match the five manual domains, and research topics need not form uniformly dense clusters.

A grid search over HDBSCAN hyperparameters $\min_\text{cluster\_size} \in \{5, 10, 15, 20, 25\}$ and $\min_\text{samples} \in \{1, 3, 5, 10\}$ was conducted and evaluated by silhouette score on non-noise points, selecting the configuration that maximized clustering silhouette. The selected configuration ($\min_\text{cluster\_size} = 20$, $\min_\text{samples} = 10$) produced four semantically coherent clusters with 8 noise points and a clustering silhouette score of 0.6852. Domain labels were not used at any stage of the clustering pipeline; they were applied only post-hoc to compute ARI~(Adjusted Rand Index) and NMI~(Normalized Mutual Information) as measures of agreement between the discovered clusters and manual domain assignments.

As a robustness check, the analysis was repeated using library-default UMAP parameters ($n_\text{neighbors} = 15$, $\min_\text{dist} = 0.1$) with the same HDBSCAN configuration, to verify that the clustering structure reflects genuine semantic organization in the corpus rather than an artifact of hyperparameter optimization.

\subsection*{Statistics}
Non-parametric tests were used throughout owing to non-normal citation distributions. Differences in citation counts across domains were assessed
using the Kruskal-Wallis test with post-hoc Dunn's test (Bonferroni-corrected).  The association between country-level publication volume and median citation count was tested by Spearman rank correlation, which makes no normality assumption and is robust to the outliers that characterise citation data; this analysis was restricted to countries with $\geq 10$ first-authored papers to ensure that median citation counts are based on a sufficient number of papers to be meaningful.
A chi-square test was used to assess independence between thematic domain and first author's country of affiliation, as both variables are categorical; 
this test was restricted to the same set of countries with $\geq 10$ first-authored papers used for the Spearman analysis (11 countries $\times$ 5 domains), to avoid sparse country-domain cells.

Cram'{e}r's $V$ was reported alongside it since the chi-square statistic grows with sample size and cannot be used to compare across studies or interpret whether the association is weak or strong. 
Geographic and institutional concentration were measured with the Gini coefficient, which captures how unevenly a quantity is distributed across a population - in this context, how research output is distributed across countries or institutions, from 0 (perfect equality) to 1 (all output in a single entity).
Bootstrap confidence intervals (10,000 resamples) were computed for all reported median citation counts, as the skewness of citation distributions makes parametric confidence intervals unreliable. 
A bootstrap CI was also computed for the geographic Gini coefficient. Since no standard formula exists to compute a CI for the Gini directly, the corpus papers were resampled with replacement (10{,}000 iterations). The Gini was recomputed from each resample, and the 2.5th and 97.5th percentiles were taken as the 95\% CI.

Two silhouette scores are reported to distinguish two separate aspects of the clustering pipeline. The \textit{embedding silhouette} is computed on the two-dimensional UMAP projection using manual domain   
labels, and quantifies how well the projection separates the five thematic domains (range $-1$ to $+1$; higher is better), which we used as the  
model and hyperparameter selection criterion. The \textit{clustering silhouette} is computed on HDBSCAN cluster assignments~(excluding noise     
points), and measures internal cluster cohesion independently of domain labels. ARI and NMI
measure agreement between HDBSCAN cluster assignments and manual domain labels~(0 = no better than chance; 1.0 = perfect agreement).

\subsection*{Use of AI writing tools}

AI-assisted writing tools were used to improve the clarity and fluency of the manuscript. All scientific content, data analysis, interpretations, and conclusions are the authors' own. The authors take full responsibility for the integrity of the work.

\newpage

%%=============================================================%%
%% DATA AVAILABILITY
%%=============================================================%%
 
\section*{Data availability}
 
% The corpus metadata and \textit{interactive dashboard} are publicly available
% at \url{https://biasatlas.cair-nepal.org}. Source
% code for the dashboard is available at the same repository. Corpus
% metadata was retrieved from the OpenAlex API
% (\url{https://openalex.org}).

The corpus metadata, \textit{interactive dashboard}, and accompanying dataset are publicly available at
\url{https://biasatlas.cair-nepal.org}, while the complete bibliographic dataset can be accessed on Hugging Face at
\url{https://huggingface.co/datasets/cair-nepal/ai-bias-research-landscape}.

%%=============================================================%%
%% CODE AVAILABILITY
%%=============================================================%%
 
\section*{Code availability}
 
Code used for bibliometric analysis, semantic clustering, and figure. The 
generation is available at the project repository
(\url{https://github.com/CAIRNepal/biasatlas}).
 
\section*{Author contributions}

\textbf{Abhash Shrestha} led the implementation of the analytical workflow, performed the bibliometric, geographic, citation and thematic analyses, interpreted the results and wrote the initial manuscript draft.  
\textbf{Subigya Gautam} contributed to data analysis, corpus curation, thematic classification, interpretation of results and manuscript writing. \textbf{Anu Sapkota} contributed to corpus construction, record screening, thematic classification, data validation and manuscript revision. \textbf{Sanju Tiwari} contributed to methodological guidance, interpretation of findings, manuscript review and revision. \textbf{Tek Raj Chhetri} conceived the study, developed the central research idea, designed the research framework, supervised the study, provided methodological guidance, guided the implementation and analysis, and contributed to manuscript writing and revision.

%%=============================================================%%
%% COMPETING INTERESTS
%%=============================================================%%
 
\section*{Competing interests}
 
The authors declare no competing interests.

% ============================================================
% REFERENCES (placeholder - use your .bib file)
% ============================================================
% \bibliographystyle{sn-mathphys-num}
\bibliography{sn-bibliography}

\end{document}